\documentclass[aps,prl,nobibnotes,twocolumn,showkeys,superscriptaddress]{revtex4-1}

\usepackage[artemisia]{textgreek}
\usepackage{amssymb}
\usepackage{epsfig}
\usepackage{amsmath}
\usepackage{graphicx}
\usepackage{dcolumn}
\usepackage{latexsym}
\usepackage{color}
\usepackage{epstopdf}
\usepackage{subfigure}
\usepackage{nicefrac}
\usepackage{blindtext}
\usepackage[ulem=normalem]{changes}
\usepackage{float}
\usepackage{xcolor}
\usepackage{soul}
\usepackage[T1]{fontenc}
\usepackage[hidelinks]{hyperref}
\usepackage{xcolor}
\setcitestyle{super}
\hypersetup{
    colorlinks,
    linkcolor={red!50!black},
    citecolor={blue!50!black},
    urlcolor={blue!80!black}
}
\usepackage{lipsum}
\usepackage[utf8]{inputenc}
\usepackage{pdfpages}
\usepackage{pgffor}

\makeatletter
\AtBeginDocument{\let\LS@rot\@undefined}
\makeatother

\def\supplementfilename{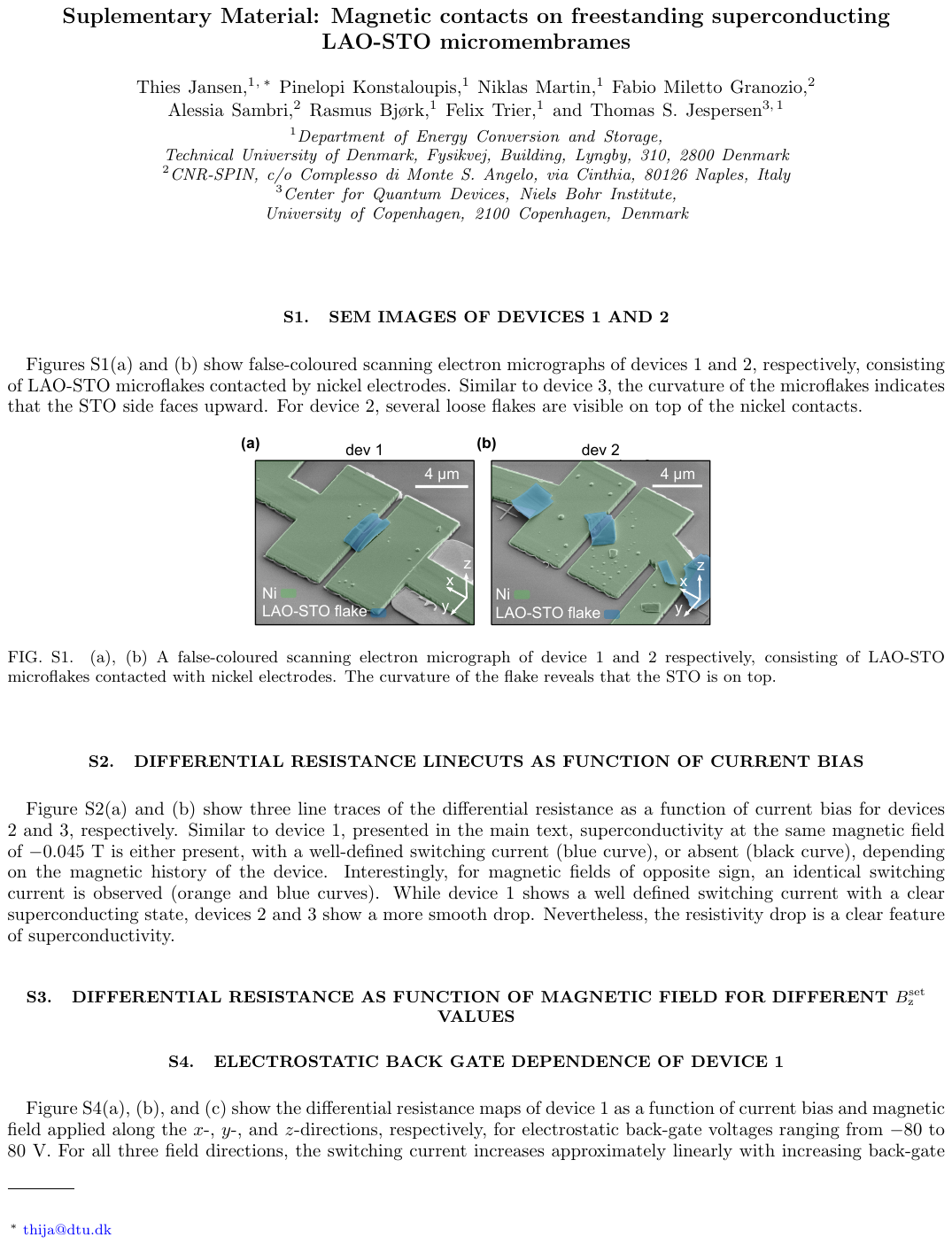}

\pdfximage{\supplementfilename}
\def\numbersupplementpages{\the\pdflastximagepages}

\usepackage[symbol]{footmisc}

\begin{document}

\title{Magnetic contacts on freestanding superconducting LaAlO$_3$/SrTiO$_3$ micromembranes}

\author{Thies Jansen}
\email{thija@dtu.dk}
\affiliation{Department of Energy Conversion and Storage, Technical University of Denmark, Fysikvej, Lyngby, 310, 2800 Denmark}

\author{Pinelopi Konstantinopoulou}
\affiliation{Department of Energy Conversion and Storage, Technical University of Denmark, Fysikvej, Lyngby, 310, 2800 Denmark}

\author{Niklas Martin}
\affiliation{Department of Energy Conversion and Storage, Technical University of Denmark, Fysikvej, Lyngby, 310, 2800 Denmark}

\author{Fabio Miletto Granozio}
\affiliation{CNR-SPIN, c/o Complesso di Monte S. Angelo, via Cinthia, 80126 Naples, Italy}

\author{Alessia Sambri}
\affiliation{CNR-SPIN, c/o Complesso di Monte S. Angelo, via Cinthia, 80126 Naples, Italy}

\author{Rasmus Bjørk}
\affiliation{Department of Energy Conversion and Storage, Technical University of Denmark, Fysikvej, Lyngby, 310, 2800 Denmark}

\author{Felix Trier}
\affiliation{Department of Energy Conversion and Storage, Technical University of Denmark, Fysikvej, Lyngby, 310, 2800 Denmark}

\author{Thomas S. Jespersen}

\affiliation{Center for Quantum Devices, Niels Bohr Institute, University of Copenhagen, 2100 Copenhagen, Denmark}
\affiliation{Department of Energy Conversion and Storage, Technical University of Denmark, Fysikvej, Building, Lyngby, 310, 2800 Denmark}

\date{\today}

\begin{abstract}

The superconducting two-dimensional electron gas (2DEG) at the LaAlO$_3$/SrTiO$_3$ (LAO-STO) interface is a promising platform for superconducting spintronics, however, integrating ferromagnetic contacts with the superconducting 2DEG remains challenging. Here, we realize superconducting LAO-STO micromembrane devices contacted by ferromagnetic nickel contacts through a side-contact geometry. Low-temperature transport measurements demonstrate that superconductivity is preserved in the presence of the ferromagnetic contacts. We show that the superconducting state is strongly influenced by the magnetic history of the nickel contacts, which generates a tunable effective magnetic field in the 2DEG. Through an effective field model, the magnetization of the contacts can be inferred from the maximum superconducting response. Our results establish ferromagnetically contacted LAO-STO as a platform for future investigations of spin injection into oxide superconductors and provide a route towards superconducting spintronic devices based on complex oxide interfaces.

\end{abstract}


\maketitle

\section{Introduction}
The interplay between ferromagnetism and superconductivity is of considerable interest for superconducting memory applications \cite{caruso_properties_2018, fermin_mesoscopic_2022, caruso_rf_2018}, which exploit the sensitivity of the superconducting critical current to the non-volatile magnetic state of nearby ferromagnets. In addition, the coexistence of superconductivity and magnetism can give rise to novel superconducting states, such as spin-triplet superconductivity \cite{bergeret_odd_2005, buzdin_proximity_2005}. This phenomenon is exploited in device geometries such as ferromagnet–superconductor–ferromagnet junctions, where the spin-triplet supercurrent depends on the relative magnetization orientation of the two ferromagnetic contacts, allowing the junction to function as a controllable superconducting spin valve \cite{visani_equal-spin_2012, cai_evidence_2021, li_superconducting_2013, wang_giant_2014}. Such devices form a promising basis for superconducting spintronic applications \cite{linder_superconducting_2015}.

In these systems, the magnetic material is typically embedded in a planar heterostructure, and transport occurs either through the unpatterned heterostructure \cite{lesne_highly_2016} or across a planar junction \cite{cai_evidence_2021, visani_equal-spin_2012, wang_giant_2014, strambini_superconducting_2022, yang_extremely_2010}. This geometry offers several advantages. Due to shape anisotropy, the magnetization is predominantly uniform and in-plane, while the magnetic layers can be made only a few nanometers thick, thereby minimizing stray magnetic fields that could otherwise affect the superconducting state. However, this approach largely limits the material platform to conventional superconducting thin films.

\begin{figure*}
    \centering
    \includegraphics[scale=0.9]{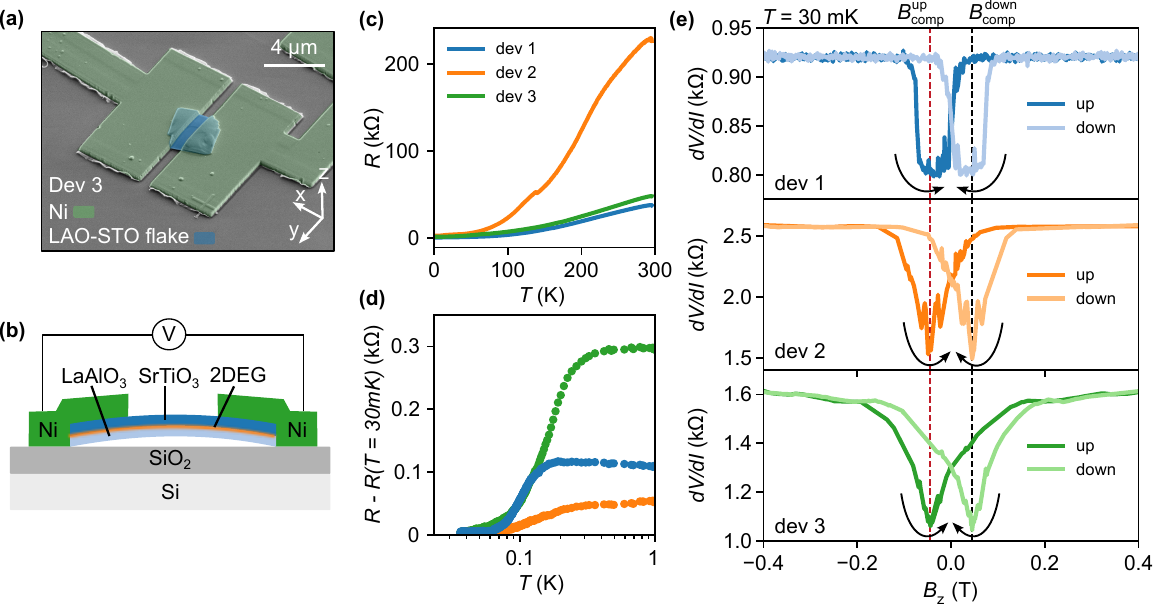}
    \caption{(a) A false-coloured scanning electron micrograph of a LAO-STO micromembrane contacted with nickel  contacts. The curvature of the micromembrane reveals that the STO is on top. (b) schematic of the device. (c) The two point resistance as function of the temperature from 300 K to 1 K for three different devices. Here the residual resistance at 30 mK, including the contact and filter resistance is subtracted. (d) The two point resistance as function of temperature from 1 K down to 30 mK revealing the superconducting state in all three devices. (e) Differential resistance as function of the out-of-plane magnetic field for positive and negative field sweep directions, displaying different resistance minima. The measured resistance includes the contact resistance.}
    \label{fig:figure1}
\end{figure*}
The two-dimensional electron gas (2DEG) and the accompanying superconducting state that emerge at the interface between the two oxide insulators LaAlO$_3$ and SrTiO$_3$ (LAO-STO) \cite{ohtomo_high-mobility_2004, reyren_superconducting_2007, chen_laalo3srtio3_2024} provide an attractive alternative platform for superconducting spintronics. In addition to superconductivity, the LAO-STO interface exhibits gate-tunable Rashba spin–orbit coupling (SOC) \cite{caviglia_tunable_2010, niu_giant_2017, diez_giant_2015, benshalom:2010, vaz_determining_2020}, gate control of the superconducting state, and coexistence with interfacial magnetism \cite{brinkman_magnetic_2007}. Furthermore, spin-polarized currents have already been injected into the LAO-STO 2DEG using ferromagnetic contacts, revealing efficient spin-to-charge conversion \cite{vaz_mapping_2019, lesne_highly_2016, song_observation_2017}. However, these studies have so far been restricted to the normal state.

The influence of ferromagnetic contacts on the superconducting LAO-STO 2DEG remains largely unexplored, because extending such experiments into the superconducting regime is challenging. The superconducting 2DEG is buried below the surface, making the fabrication of low-resistance ferromagnetic contacts difficult. Moreover, ferromagnetism generally tends to suppress superconductivity, and the device geometries required to contact the buried 2DEG often involve larger magnetic volumes than those used in conventional planar heterostructures. As a result, stray magnetic fields generated by the ferromagnetic contacts can become an important factor in determining the superconducting properties of the device. 

The recent development of synthesis and transfer techniques for freestanding complex-oxide membranes offers a route to address these challenges while providing the design flexibility of conventional semiconductor platforms \cite{lu:2016, lu:2019, sambri:2020, kum:2020, eom:2021, pryds:2024}. In particular, spalled LAO-STO micromembranes retain superconductivity \cite{erlandsen:2022}, while their reduced lateral dimensions and exposed sidewalls provide access to the buried LAO-STO interface from the side. This enables device geometries that are difficult to realize in conventional planar LAO-STO heterostructures and provides a route to integrate ferromagnetic contacts directly with the superconducting 2DEG.

Here, we use LAO-STO micromembranes transferred onto silicon substrates and contact the 2DEG from the side with nickel contacts, similar to the work on carbon nanotubes devices \cite{hauptmann_electric-field-controlled_2008, gaass_universality_2011}. Through low-temperature transport measurements, we demonstrate the coexistence of superconductivity and ferromagnetism in these devices and show that the superconducting state can be controlled through the magnetic history of the ferromagnetic contacts. Furthermore, we argue that the sensitivity of the superconducting 2DEG can be used as an internal magnetometer of the nickel contacts and demonstrate that we can engineer an anisotropic hysteretic response on the superconducting state by the applied field. Our results establish a route towards ferromagnet–superconductor devices based on LAO-STO and provide a platform for future studies of spin injection, triplet superconductivity, and superconducting spintronic functionality.

\section{Results}
LAO-STO micromembranes were released by strain induced spalling \cite{sambri:2020} and are mechanically transferred onto a Si/SiO$_2$ device substrate. The 2DEG at the LAO-STO interface is contacted using electron-beam lithography, followed by a multi-step angled evaporation of Ni, similar to the method reported in \cite{meucci_intrinsic_2026} (see methods for details). Figure \ref{fig:figure1}(a) shows a false-coloured scanning electron micrograph of a typical device. In total three devices are measured. Due to internal strain arising from the lattice mismatch between LAO and STO, the micromembrane is slightly curved. The curvature of device 3 in figure \ref{fig:figure1}(a) shows that the micromembrane is oriented with the STO side facing upwards, as schematically illustrated in figure \ref{fig:figure1}(b). Device 1 and 2 have the same orientation (see supplementary material section S1). The STO and LAO layer are approximately 70 nm thick and ohmic electrical contact between the nickel and the 2DEG occurs from the side. 

Figure \ref{fig:figure1}(c) shows the two-point longitudinal resistance as a function of temperature for three devices, all exhibiting metallic behaviour. A zoom of the same dataset below 1 K is presented in figure \ref{fig:figure1}(d), where a superconducting transition is observed in all three devices between 100 and 200 mK.  

To investigate the interaction between the magnetic state of the nickel contacts and the superconducting LAO-STO 2DEG in the micromembrane, we measure the differential resistance as a function of out-of-plane magnetic field, $B_\mathrm{z}$, for all three devices. The results are shown in figure \ref{fig:figure1}(e). For all devices, a  magnetic field of approximately 100 mT suppresses the superconducting state, comparable to the critical out-of-plane field for LAO-STO micromembranes previously reported \cite{erlandsen:2022}. Moreover, a clear difference is observed between upward and downward magnetic field sweeps. This hysteresis indicates the presence of magnetic order in the device and highlights the sensitivity of the superconducting state to the magnetic state of the nickel contacts.

Notably, the resistance minimum does not occur at zero magnetic field but instead at finite fields of $\pm 45$ mT for all three devices, as indicated by the dashed line in figure \ref{fig:figure1}(e). We interpret this as the value of the external field, $B_\mathrm{comp}^{\mathrm{up},\mathrm{down}}$, which compensates for the field from the Ni contacts, and thus minimizes the field in the 2DEG, $B_\mathrm{2DEG}$.

This also explains why the decrease in resistance upon lowering the temperature with $B_\mathrm{z} = 0$, as shown in figure \ref{fig:figure1}(d), is less pronounced than the resistance variation observed as a function of magnetic field in figure \ref{fig:figure1}(e). 

\begin{figure*}
    \centering
    \includegraphics[scale=0.9]{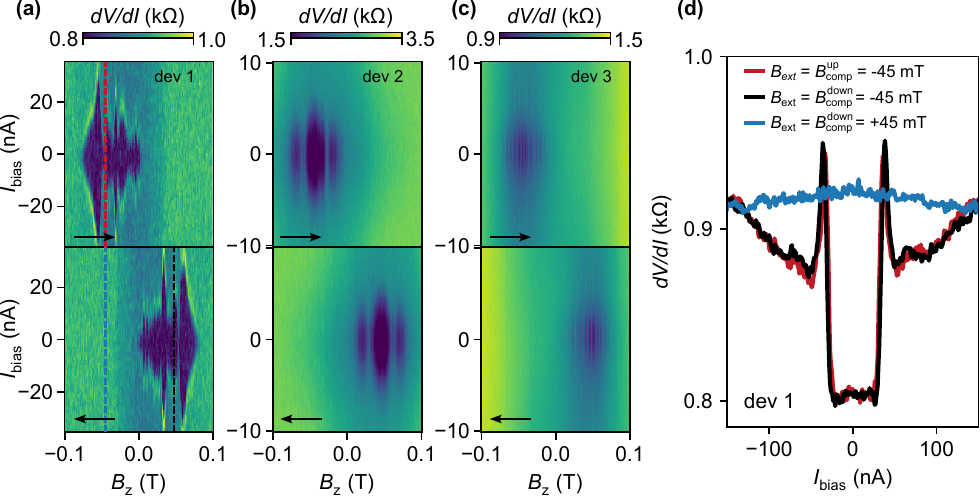}
    \caption{(a-c) Differential resistance as function of current bias and magnetic field for the positive and negative sweeping direction for three different devices. The differential resistance maps reflect the symmetry $R(I_{\mathrm{bias}},B_\mathrm{z})=R(I_\mathrm{bias},-B_\mathrm{z})$  (d) Differential resistance of device 1 as function of current bias and external magnetic field for three magnetic field sweeping histories. The blue and black curve show that drastically different superconducting states can be achieved at the same field, but with different sweeping directions. The blue and orange curve show that the same switching current is observed at different magnetic fields.}
    \label{fig:figure2A}
\end{figure*}

The superconducting state at different fields is characterized in more detail by measuring the differential resistance map as function of current bias and $B_\mathrm{z}$ in both sweeping directions for all three devices, as shown in figure \ref{fig:figure2A}(a-c). For all three devices, the maps show again a clear hysteretic dependence on the sweeping direction and that the highest switching current during the sweep occurs at $B_\mathrm{comp}^{\mathrm{up},\mathrm{down}}$ (depending on the sweep direction), before the field reaches zero. 


Figure \ref{fig:figure2A}(d) shows three line traces of the differential resistance as function of current bias for device 1. For the same magnetic field of -45 mT, superconductivity can be present with a clear switching current (blue curve) or is absent (black curve), depending on the magnetic history. Interestingly, at fields of opposite sign, an identical switching current is observed (orange and blue curve). 

Devices 2 and 3 exhibit the same qualitative behavior as presented in S2, although the switching current is less sharply defined. In addition, both devices show modulation of $I_c$ as function $B_\mathrm{z}$. We attribute these modulations to the intrinsic inhomogeneities in the LAO-STO 2DEG, as reported previously in LAO-STO micromembranes with non-magnetic metalic leads \cite{erlandsen:2022, noteGunjan}.
 
\begin{figure*}
    \centering
    \includegraphics[scale=0.9]{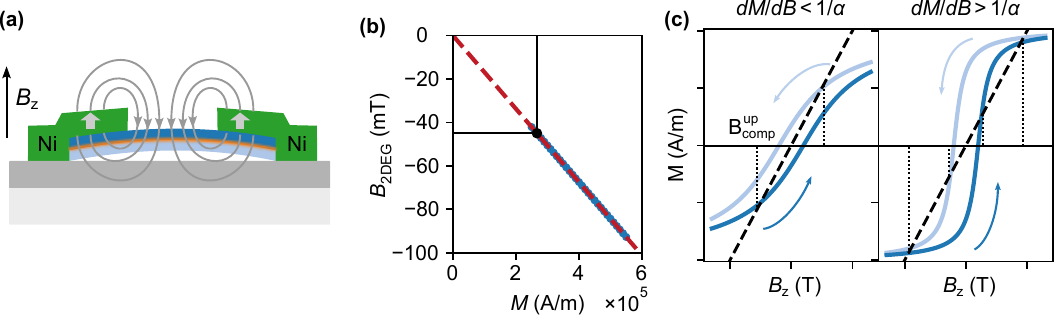}
    \caption{(a) Schematic displaying the field lines for magnetized contacts in the positive z-direction. (b) Results of the finite element simulations relating the the effective field in the 2DEG at $B_{\mathrm{z}} = 0$ to the magnetization of the contacts. (c) The condition that $B_{\mathrm{2DEG}}(B_\mathrm{z}) = 0$, is qualitatively obtained by the crossings of $M_\mathrm{z}$ with the dashed line $B_\mathrm{z}/\alpha$. Depending on the steepness and remanence of $M_\mathrm{z}(B_\mathrm{z})$, $B_{\mathrm{2DEG}}(B_\mathrm{z})$ vanishes once or three times for a single sweep direction.}
    \label{fig:figure2B}
\end{figure*}

The hysteresis in the superconducting state can be attributed to the combined effect of the external magnetic field and the the magnetic field generated by the magnetized contacts. For nickel contacts fully polarized in the positive z-direction, a magnetic field is generated that opposes the externally applied field within the superconducting 2DEG. The situation is illustrated schematically in figure \ref{fig:figure2B}(a). As a consequence, the effective magnetic field at the 2DEG, $B_\mathrm{2DEG}$, is minimum 
at a finite positive value for the compensating external field $B_{\mathrm{comp}}$. Conversely, when the contacts are magnetized in the negative z-direction, the compensation field occurs at a finite negative applied field. 

Since the nickel contacts are micrometers in size, in the intermediate magnetization regime the nickel contacts are expected to consist of multiple magnetic domains. Although shape anisotropy favors an in-plane magnetization, the domain walls separating neighboring domains generate localized out-of-plane stray magnetic fields that can suppress superconductivity \cite{bhatia_domain_2019, aladyshkin_reverse-domain_2010, gillijns_domain-wall_2005}. Consequently, during magnetization reversal the superconducting 2DEG experiences an inhomogeneous effective magnetic field, making its transport properties highly sensitive to the magnetic domain configuration. Experimentally, this behavior is consistent with the irregular switching current as function of $B_\mathrm{z}$ observed for device 1 in figure \ref{fig:figure2A}(a).


Finite-element simulations incorporating the geometry of the nickel contacts (see Supplementary material Section S5) provide the basis for an effective-field description. By calculating the average stray field in the 2DEG as a function of the magnetization of the contacts, we find a linear relation between the contact magnetization and the resulting field correction, as shown in figure \ref{fig:figure2B}(b) for $B_\mathrm{z}=0$. This allows the effective field experienced by the 2DEG to be expressed as
$B_{\mathrm{2DEG}}(B_\mathrm{z}) = B_\mathrm{z} - \alpha M_\mathrm{z}(B_\mathrm{z})$, $M_\mathrm{z}(B_\mathrm{z})$ is the magnetization of the contacts along the $z$-direction, and $\alpha$ is a proportionality constant determined by the device geometry. From the slope of the simulated relation in figure \ref{fig:figure2B}(b), we obtain $\alpha = 1.69\times10^{-7}$ H/m.

The behavior of $B_{\mathrm{2DEG}}$ is therefore determined by, $M_\mathrm{z}(B_\mathrm{z})$. Figure \ref{fig:figure2B}(c) schematically shows, $M_\mathrm{z}(B_\mathrm{z})$ as a function of the externally applied magnetic field, for two representative regimes. The condition for which the external field compensates the field from the nickel leads $B_{\mathrm{2DEG}}(B_\mathrm{comp}^{\mathrm{up},\mathrm{down}}) = 0$, is qualitatively obtained by the crossings of the linear dashed line representing $M_\mathrm{z} = B_\mathrm{z}/\alpha$ with the magnetization curves.

For a small coercive field, depending on the steepness of $M_\mathrm{z}(B_\mathrm{z})$, $B_{\mathrm{2DEG}}(B_\mathrm{z})$ vanishes (during a single up or down sweep) a single time ($dM/dB < 1/\alpha$) or three times ($dM/dB > 1/\alpha$).
The measurements in figure \ref{fig:figure1}(e) show a single $B_\mathrm{comp}^{\mathrm{up},\mathrm{down}}$, suggesting that nickel contacts correspond to the $dM/dB < 1/\alpha$ regime and implies some magnetic hysteresis in the nickel contacts.

\begin{figure}
    \centering
    \includegraphics[width=\linewidth]{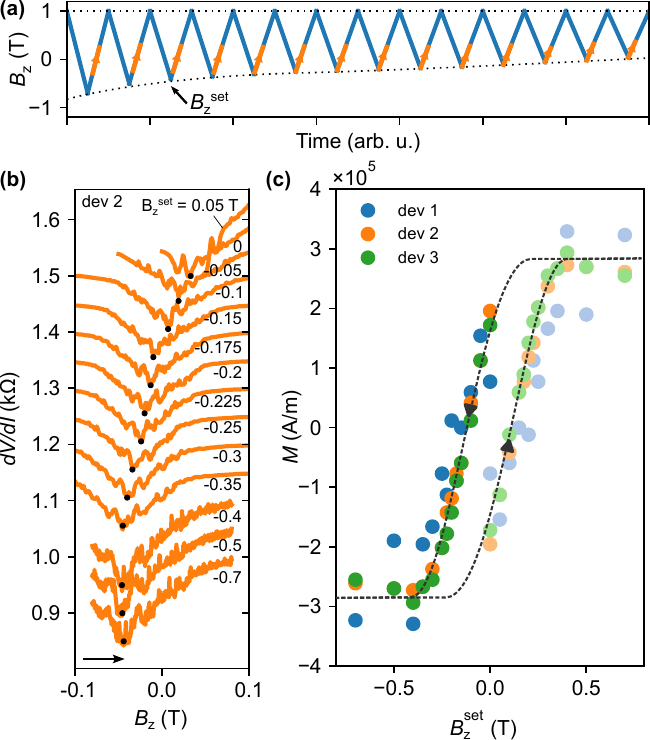}
    \caption{By avoiding complete magnetization reversal of the nickel contacts, the effective magnetic field experienced by the 2DEG can be continuously tuned. (a) shows schematically the applied magnetic field as function of time. The applied field is set at 1 T to initialize the magnetization and subsequently swept to a value $B_{\mathrm{z}}^{\mathrm{set}} < 1$ T to partially reverse the magnetization. On the back sweep, the traces in (b) are measured, as indicated by the orange marked lines. (c) Effective magnetization within our effective field model of the nickel contacts, $M_{\mathrm{z}}$, as a function of $B_{\mathrm{z}}^{\mathrm{set}}$.}
    \label{fig:figure3}
\end{figure}

Based on the finite element simulation of \ref{fig:figure3}(b), the observed compensation field of $\pm 45$ mT, corresponds to a magnetization of $2.5\times10^5$ A/m. Saturation magnetization values of nickel thin films reported in literature range from $2.3 \times 10 ^5$ - $3\times 10 ^5$ A/m   \cite{potocnik_exploring_2024, poddar_magnetic_2024}, which is in good agreement with our observations and suggest that at a field of of $\pm 45$ mT, the contacts are still in the saturation magnetization regime. 

In the measurements in figures \ref{fig:figure1} and \ref{fig:figure2A} the magnetic contacts have been swept through their full magnetization loop, resulting in a fully magnetized state. By controlling the magnetization of the contacts, i.e., by avoiding complete magnetization reversal, the effective magnetic field experienced by the device can be tuned and the magnetization of the nickel contacts can be estimated. 

To control the magnetization, the contacts are first magnetized in the positive $z$-direction by applying an external magnetic field of $B_\mathrm{z} = 1$ T. Subsequently, the field is swept to a negative value $B_{\mathrm{z}}^{\mathrm{set}}$ T, thereby (partially) magnetizing the contacts in the opposite direction. Finally, $B_\mathrm{z}$ is swept to 1 T, while measuring the differential resistance. This procedure is schematically illustrated in figure \ref{fig:figure3}(a) where the applied magnetic field is plotted as function of time .

Figure \ref{fig:figure3}(b) shows a series of differential resistance measurements for device 2 as a function of magnetic field for different values of $B_{\mathrm{z}}^{\mathrm{set}}$, ranging from $-0.7$ T to $0.05$ T. The black dots indicate the external compensation field, determined from a Gaussian fitting procedure. These measurements demonstrate the dependence of the compensation field on the magnetization of the nickel leads, which can be set by varying $B_{\mathrm{z}}^{\mathrm{set}}$. Similar behavior is observed for device 1 and 3 and is presented in the S3. 


If the change in magnetization when moving from $B_{\mathrm{z}}^{\mathrm{set}}$ to $B_\mathrm{comp}$ is neglible. Then, within the effective field model, the condition $B_{\mathrm{2DEG}} = 0$ at the $B_\mathrm{comp}$ directly relates the magnetization of the contacts to the measured compensation field through $M_\mathrm{z} = B_{\mathrm{comp}}/\alpha$. Using this relation, the magnetization can be reconstructed as a function of $B_{\mathrm{z}}^{\mathrm{set}}$, as shown in figure \ref{fig:figure3}(c). The negative branch is obtained directly from the extracted values of $B_\mathrm{comp}$ in figure \ref{fig:figure3}(b), while the positive branch is reconstructed using the symmetry between the upward and downward sweep directions. The dashed line serves as a guide to the eye and represents the inferred magnetization loop of the nickel contacts.

For nickel thin films, one would not necessarily expect a remanent magnetization along the $z$-direction due to the unfavorable shape anisotropy of thin films and the fact that bulk nickel is a soft ferromagnet. However, the magnetization reconstructed in figure \ref{fig:figure3}(c) should not be interpreted as the true magnetization of the nickel contacts, but rather as the effective magnetization experienced by the superconducting 2DEG within the effective field model. In particular, for $B_{\mathrm{z}}^{\mathrm{set}} = 0$ T, the contacts are not expected to remain uniformly magnetized along the positive $z$-direction. Instead, they are expected to relax into a multidomain state with predominantly in-plane magnetization. The domain walls separating these magnetic domains generate localized out-of-plane stray magnetic fields that suppress superconductivity \cite{bhatia_domain_2019, aladyshkin_reverse-domain_2010, gillijns_domain-wall_2005}. Consequently, the superconducting 2DEG experiences an effective out-of-plane magnetic field that is reflected in the reconstructed "effective" magnetization shown in figure \ref{fig:figure3}(c).





\begin{figure*}
    \centering
    \includegraphics[scale=0.9]{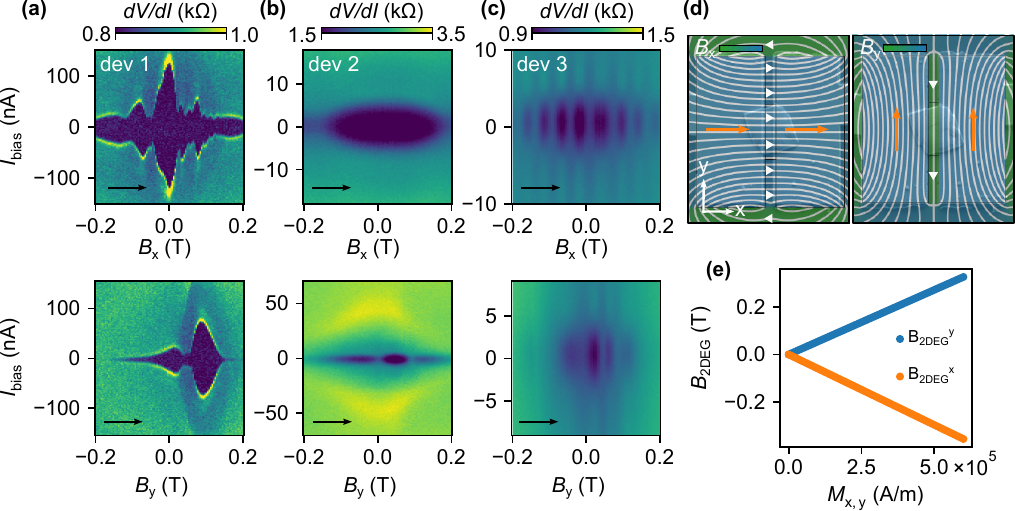}
    \caption{(a-c) Differential resistance as function of bias and magnetic field in the x and y direction for three devices. The qualitative difference between the symmetric Bx sweeps and asymmetric By sweeps can be explained by the difference in sign of the effective field experienced by the 2DEG with respect to the magnetization as illustrated in (d). (e) Finite element simulation of the effective field in the x and y direction with a different sign as function of the magnetization in the x and y direction respectively.}
    \label{fig:figure4}
 \end{figure*}

Another degree of freedom for tuning the magnetization is provided by applying a magnetic field parallel to the sample plane and the nickel contacts. In-plane magnetized contacts, and the corresponding remanent in-plane magnetic field, are expected to affect the superconducting state of the 2DEG differently. Specifically, the two-dimensional nature of the superconducting 2DEG suppresses orbital pair breaking for an in-plane magnetic field, leading to an expected larger in-plane critical field than for a perpendicular field \cite{, kim_intrinsic_2012, kozuka_two-dimensional_2009}. Moreover, the geometry of the nickel contacts favors an in-plane magnetization due to shape anisotropy, which will lead to behavior distinct from that observed for out-of-plane magnetic fields.

Figure \ref{fig:figure4}(a-c) shows the differential resistance as a function of in-plane magnetic field, $B_\mathrm{x}$ and $B_\mathrm{y}$, and current bias for all three devices. Device 1 again exhibits a well-defined switching current, with a maximum value of approximately 100 nA. This is about five times larger than the maximum switching current observed during out-of-plane field sweeps. We attribute this enhancement to the fact that the in-plane magnetic field magnetizes the contacts within the sample plane more uniformly, resulting in a negligible out-of-plane stray field experienced by the 2DEG, also at zero field. 

The difference in maximum switching current between the out-of-plane and in-plane field sweeps provides further evidence that the contacts do not necessarily relax into a purely uniform in-plane magnetized state at zero applied field. Instead, they support the presence of a multidomain state with a significant effective out-of-plane magnetization component after sweeping the out-of-plane magnetic field, consistent with the effective field model discussed above.

Another notable observation is the dependence of the compensation field on the sweep direction. For out-of-plane field sweeps ($B_\mathrm{z}$), the compensation field shifts to negative (positive) values for positive (negative) sweep directions. In contrast, for $B_\mathrm{y}$ sweeps the shift has the opposite sign, while for $B_\mathrm{x}$ sweeps the compensation field remains close to zero.

This behavior can be understood from the geometry of the devices. In all three devices, the gap between the nickel contacts is oriented along the $y$-direction, as shown in figure \ref{fig:figure1}(a) and figure S1. Consequently, the $y$-axis forms the magnetic easy axis of the contacts, giving rise to a steeper magnetization curve around the coercive field than for magnetization along the $x$-axis due to shape anisotropy. As a result, during $B_\mathrm{y}$ sweeps the condition $dM/dB > 1/\alpha$ can be satisfied, corresponding to regime 2 of the effective field model shown in figure \ref{fig:figure3}(c). The additional compensation fields predicted by this model are not resolved experimentally, which we attribute to spatial inhomogeneity of the stray field generated by the nickel contacts.

The device geometry also determines the sign of the effective field. Magnetization along the $y$-axis generates a magnetic field at the position of the 2DEG opposite to the magnetization direction, similar to the out-of-plane case, whereas magnetization along the $x$-axis produces a field in the same direction as the lead magnetization. Finite-element simulations of the magnetic field distribution and the resulting effective field, shown in figure \ref{fig:figure4}(e), confirm this behavior. Consequently, for $B_\mathrm{x}$ sweeps the compensation field is determined by the condition $B_\mathrm{x} = -\alpha M_\mathrm{x}(B_\mathrm{x})$, giving rise to an opposite field shift compared to the $B_\mathrm{z}$ and $B_\mathrm{y}$ sweeps. Because the coercive field along the $x$-direction is expected to be small due to shape anisotropy, the resulting shift remains close to zero.

These results demonstrate that the magnetic field experienced by the superconducting 2DEG is strongly determined by the geometry of the ferromagnetic contacts. Consequently, anisotropic magnetic hysteresis of the switching current can be engineered through the geometrical design of the contacts.

\section{Outlook for superconducting spin transport}

By applying an external magnetic field, the magnetization of the nickel contacts, and therefore the spin polarization of the injected electrons, can be controlled \cite{visani_equal-spin_2012, cai_evidence_2021}. Injecting spin-polarized electrons into a conventional spin-singlet superconductor suppresses Andreev reflection at the ferromagnet-superconductor interface, since the availability of electrons with opposite spin is reduced. Consequently, the observation of a supercurrent through the device requires one or more mechanisms that restore Cooper-pair transport across the interface.

Several mechanisms may contribute. First, spin-triplet Cooper pairs may be generated at the ferromagnet-superconductor interface, allowing Andreev reflection to proceed despite the spin polarization of the injected current \cite{bergeret_odd_2005, buzdin_proximity_2005}. Second,  spin relaxation within the 2DEG and at the interface may reduce the spin polarization before the electrons reach the superconducting region, effectively restoring conventional spin-singlet pairing. Finally, the strong Rashba spin-orbit coupling of the LAO-STO interface may induce spin precession and singlet-triplet conversion, thereby facilitating Cooper-pair transport.

For the present devices, Rashba-induced spin precession is expected to play a dominant role. While the spin diffusion length in the LAO-STO 2DEG is on the order of micrometers \cite{jin_nonlocal_2017, trier_electric-field_2019, jin_pure_2021}, comparable to the device dimensions, the spin precession length associated with Rashba spin-orbit coupling is expected to be considerably shorter. As a result, the injected spins are expected to undergo substantial rotation before traversing the superconducting channel, reducing the importance of preserving their initial spin orientation and potentially facilitating singlet-triplet conversion.

A direct signature of spin-triplet superconductivity would require independent control over the magnetization of the two ferromagnetic contacts. In a ferromagnet-superconductor-ferromagnet junction, the supercurrent depends on the relative orientation of the two magnetizations, allowing triplet transport to be identified by comparing the parallel and antiparallel magnetic configurations \cite{wang_giant_2014, linder_superconducting_2015}. Since the magnetization of the two nickel contacts cannot be controlled independently in the present device geometry, such measurements are not yet possible.

The side-contact geometry employed here also provides an interesting platform for studying the interplay between Rashba spin-orbit coupling and superconductivity. Unlike conventional planar junctions, inversion symmetry is broken both at the LAO-STO interface and locally at the side-contact interface, leading to a more complex spin texture whose influence on singlet-triplet conversion remains to be understood. Furthermore, as we demonstrate in supplementary section S4, it is possible to tune the carier density of LAO-STO micromembranes by electrostatic gating, opening additional degrees of freedom to tune the Rashba spin-orbit coupling and superconducting state.

Future devices incorporating independently switchable, shape-anisotropic ferromagnetic contacts separated by superconducting channel shorter than the spin procession length would enable direct investigations of spin injection and triplet superconductivity in the LAO-STO 2DEG. Nevertheless, the present work demonstrates that the superconducting state is highly sensitive to the magnetic history of the ferromagnetic contacts. This effect must be carefully considered when designing future superconducting spintronic devices based on LAO-STO.

\section{Conclusion}
In conclusion, we have realized superconducting LaAlO$_3$/SrTiO$_3$ micromembrane devices contacted by ferromagnetic nickel contacts through a side-contact geometry. Despite the presence of ferromagnetic contacts, all devices exhibit superconductivity at millikelvin temperatures. We show that the superconducting state is strongly influenced by the magnetic history of the nickel contacts, resulting in a hysteretic superconducting response that can be quantitatively described by an effective magnetic field generated by the ferromagnetic contacts. By controlling the magnetization of the contacts, the effective magnetic field experienced by the superconducting 2DEG can be continuously tuned, allowing the effective magnetization loop of the contacts to be reconstructed from superconducting transport measurements.

Our results establish a ferromagnet-superconductor hybrid platform based on the LAO-STO 2DEG and demonstrate that superconductivity and ferromagnetism can be integrated in this oxide interface system. This opens the door to future investigations of spin injection into the superconducting state, Rashba spin-orbit-driven superconducting transport, and spin-triplet superconductivity in gate-tunable oxide heterostructures.



\section{Methods}

\subsection{Device fabrication}

Device fabrication was performed on a $5\times5~\mathrm{mm}^2$ p-type Si(100) substrate with a $500~\mathrm{nm}$ thermally grown SiO$_2$ layer. The substrate was pre-patterned with Au contacts and a $300\times300~\mu\mathrm{m}^2$ central array of fine alignment markers. LAO-STO micromembranes were released by strain induced spalling \cite{sambri:2020} and are mechanically transferred onto the Si/SiO$_2$ device substrate to the area containing the alignment markers. An optical microscope (Nikon, bright-field, $100\times$ magnification) was used to identify suitable membranes and align the two-contact device geometry.

\textit{Electron-beam lithography.} The contact structures were defined using electron-beam lithography (EBL) with an Elionix system operated at an acceleration voltage of $125~\mathrm{kV}$ and a beam current of $1~\mathrm{nA}$. An electron-beam dose of $1200~\mu\mathrm{C}/\mathrm{cm}^2$ was used. A three-layer PMMA resist stack with a total thickness of approximately $1~\mu\mathrm{m}$ was employed. The relatively thick resist layer was chosen to accommodate the curvature of the spalled membranes and to ensure sufficient metal coverage of the sidewalls of the micromembranes, providing access to the buried LAO-STO interface. The bottom resist layer, El9, was used to create an undercut profile, which is essential for the subsequent metal lift-off process.

\textit{Argon milling and metal deposition.} To establish electrical contact to the 2DEG at the LAO-STO interface, a two-step angled Ar$^+$ milling process followed by Ti deposition was performed in an AJA deposition system. The Ar$^+$ milling was carried out at an acceleration voltage of $220~\mathrm{V}$ for $2~\mathrm{min}$ per side, with the sample oriented at an angle of $40^\circ$ relative to the surface normal, at a base pressure of approximately $10^{-8}~\mathrm{Torr}$. Ti was subsequently deposited in the same chamber without breaking vacuum using a two-step evaporation process at an angle of $20^\circ$ relative to the surface normal.

For the ferromagnetic contacts, the sample was transferred to a separate deposition system, where $120~\mathrm{nm}$ of nickel was deposited normal to the sample surface. Finally, the resist was removed by immersing the sample in 1,3-dioxolane for $1~\mathrm{h}$ at room temperature, followed by rinsing in acetone for $2~\mathrm{min}$ and isopropanol for $1~\mathrm{min}$.
         
\subsection{Magnetotransport measurements}
Magnetotransport measurements were performed in a dilution refrigerator with a base temperature of 30 mK. The sample was current-biased by applying a DC voltage bias superimposed with an AC excitation voltage at 283 Hz from a lock-in amplifier to the sample in series with a 10~M$\Omega$ (DC) and 1~G$\Omega$ (AC) resistor. The two-point voltage across the sample was amplified using a differential voltage amplifier with a gain of 1000. The AC voltage component was demodulated by the lock-in amplifier, from which the differential resistance, $dV/dI$, was obtained. The resistance of the cryostat wiring (5.8 k$\Omega$) was subtracted from the measured voltage signal.
The cryostat was equipped with a vector magnet capable of applying magnetic fields up to 6 T in the direction perpendicular to the membrane plane and current flow and up to 1 T parallel to the membrane.



\section{Acknowledgements}
T.J. and T.S.J. acknowledge support by grant no. 00070155 from the Villum Foundation. N.M., R.B. and F.T. acknowledge support by grant no. 10.46540/4254-00012B (PESTO) from Independent Research Fund Denmark. A.S. and F.M.G. acknowledge support from by grant grants no. 2022TCJP8K PRIN PNRR Foxes - and by grant no. 2022TCT72 Omega from Ministero dell’Università e della Ricerca (IT).

\section{Author Contributions}
T.J performed electrical transport measurements, analyzed the data, and wrote the initial manuscript. P.K. fabricated the devices. N.M., R.B. and T.J. performed the finite element simulations. A.S and F.M.G fabricated the micromembranes by epitaxial spalling via PLD. F.T. edited the manuscript
T.S.J conceptualized the experiment, analyzed the data and edited the manuscript.

\section{Competing financial interests}
The authors declare no competing financial interests.

\section{Supplementary material}

S1 shows extra scanning micrograph images of device 1 and 2. S2 shows the differential resistance line cuts as function of current bias for different magnetic fields for device 2 and 3. S3 shows the differential resistance as function of magnetic field for different $B_\mathrm{z}^\mathrm{set}$. S4 shows extra transport data of the differential resistance current bias field maps as function of electrostatic back gating. S5 discusses the finite element simulations in more detail.

\bibliography{ref,RefsTSJ, refTJ}

\begin{thebibliography}{45}
\bibitem [{not()}]{noteGunjan}%
  \BibitemOpen
  \href@noop {} {\enquote {\bibinfo {title} {Systemetic analysis will be presented elsewhere},}\ }\BibitemShut {NoStop}%
\end{thebibliography}

\clearpage



\section*{Supplementary material}

# Suplementary Material: Magnetic contacts on freestanding superconducting LAO-STO micromembrames

Thies Jansen,[1,*] Pinelopi Konstaloupis,[1] Niklas Martin,[1] Fabio Miletto Granozio,[2]
Alessia Sambri,[2] Rasmus Bjørk,[1] Felix Trier,[1] and Thomas S. Jespersen[3,1]
[1] *Department of Energy Conversion and Storage, Technical University of Denmark, Fysikvej, Building, Lyngby, 310, 2800 Denmark*
[2] *CNR-SPIN, c/o Complesso di Monte S. Angelo, via Cinthia, 80126 Naples, Italy*
[3] *Center for Quantum Devices, Niels Bohr Institute, University of Copenhagen, 2100 Copenhagen, Denmark*

## S1. SEM IMAGES OF DEVICES 1 AND 2

Figures S1(a) and (b) show false-coloured scanning electron micrographs of devices 1 and 2, respectively, consisting of LAO-STO microflakes contacted by nickel electrodes. Similar to device 3, the curvature of the microflakes indicates that the STO side faces upward. For device 2, several loose flakes are visible on top of the nickel contacts.

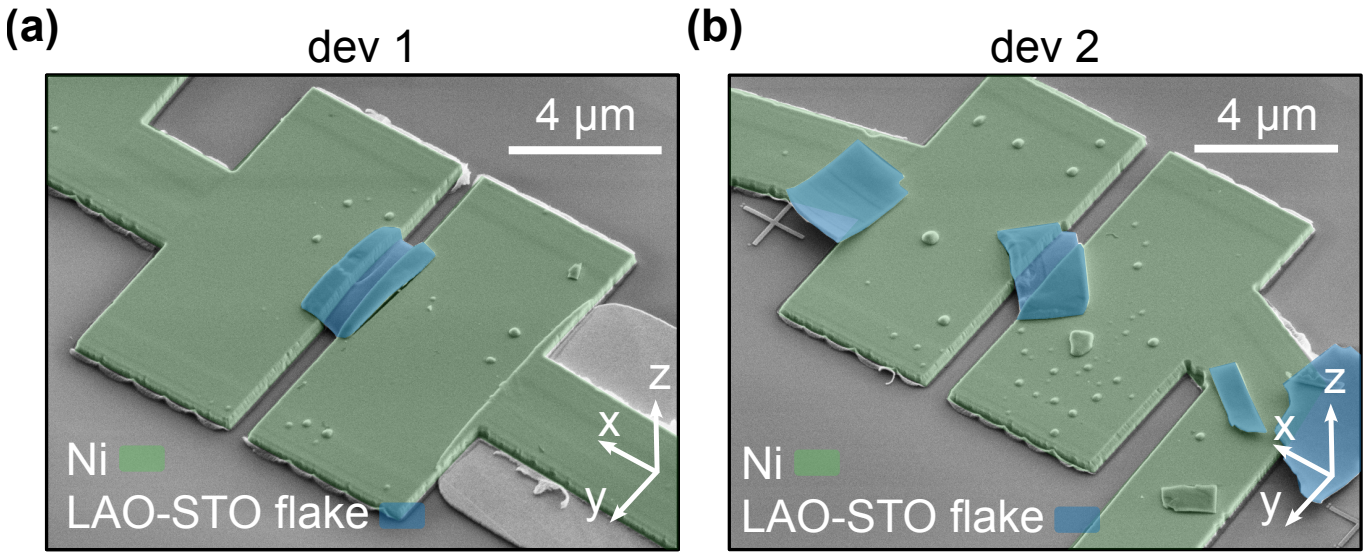


FIG. S1. (a), (b) A false-coloured scanning electron micrograph of device 1 and 2 respectively, consisting of LAO-STO microflakes contacted with nickel electrodes. The curvature of the flake reveals that the STO is on top.

## S2. DIFFERENTIAL RESISTANCE LINECUTS AS FUNCTION OF CURRENT BIAS

Figure S2(a) and (b) show three line traces of the differential resistance as a function of current bias for devices 2 and 3, respectively. Similar to device 1, presented in the main text, superconductivity at the same magnetic field of $-0.045$ T is either present, with a well-defined switching current (blue curve), or absent (black curve), depending on the magnetic history of the device. Interestingly, for magnetic fields of opposite sign, an identical switching current is observed (orange and blue curves). While device 1 shows a well defined switching current with a clear superconducting state, devices 2 and 3 show a more smooth drop. Nevertheless, the resistivity drop is a clear feature of superconductivity.

## S3. DIFFERENTIAL RESISTANCE AS FUNCTION OF MAGNETIC FIELD FOR DIFFERENT $B_z^{set}$ VALUES

## S4. ELECTROSTATIC BACK GATE DEPENDENCE OF DEVICE 1

Figure S4(a), (b), and (c) show the differential resistance maps of device 1 as a function of current bias and magnetic field applied along the $x$-, $y$-, and $z$-directions, respectively, for electrostatic back-gate voltages ranging from $-80$ to 80 V. For all three field directions, the switching current increases approximately linearly with increasing back-gate

* thija@dtu.dk

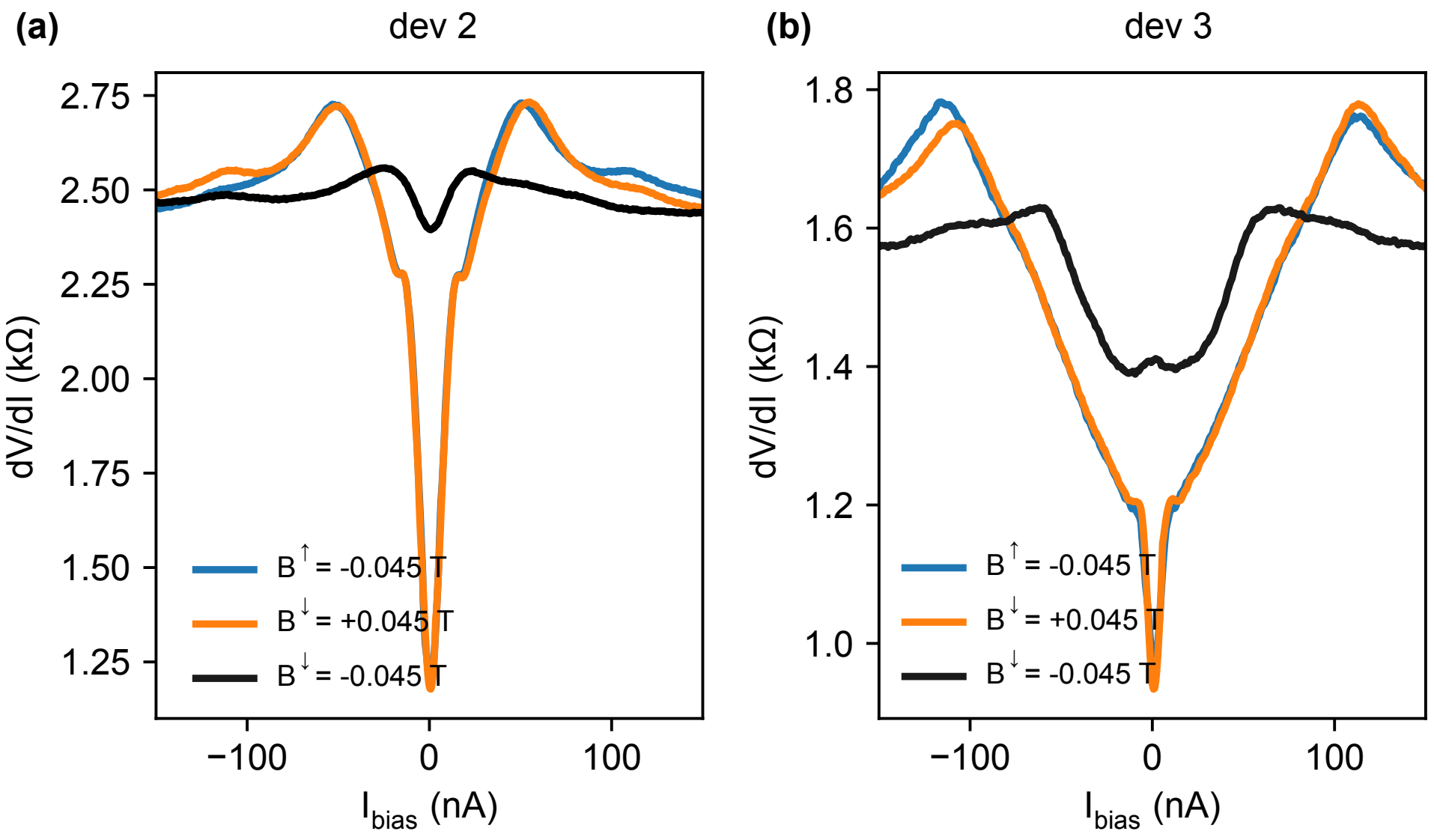


FIG. S2. Differential resistance of device 2 (a), and 3 (b), as function of current bias and external magnetic field for three magnetic field sweeping histories. Qualitatively, the same behavior is observed as device 1 presented in the main text, where a different superconducting states can be achieved at the same field depending on the magnetic history.

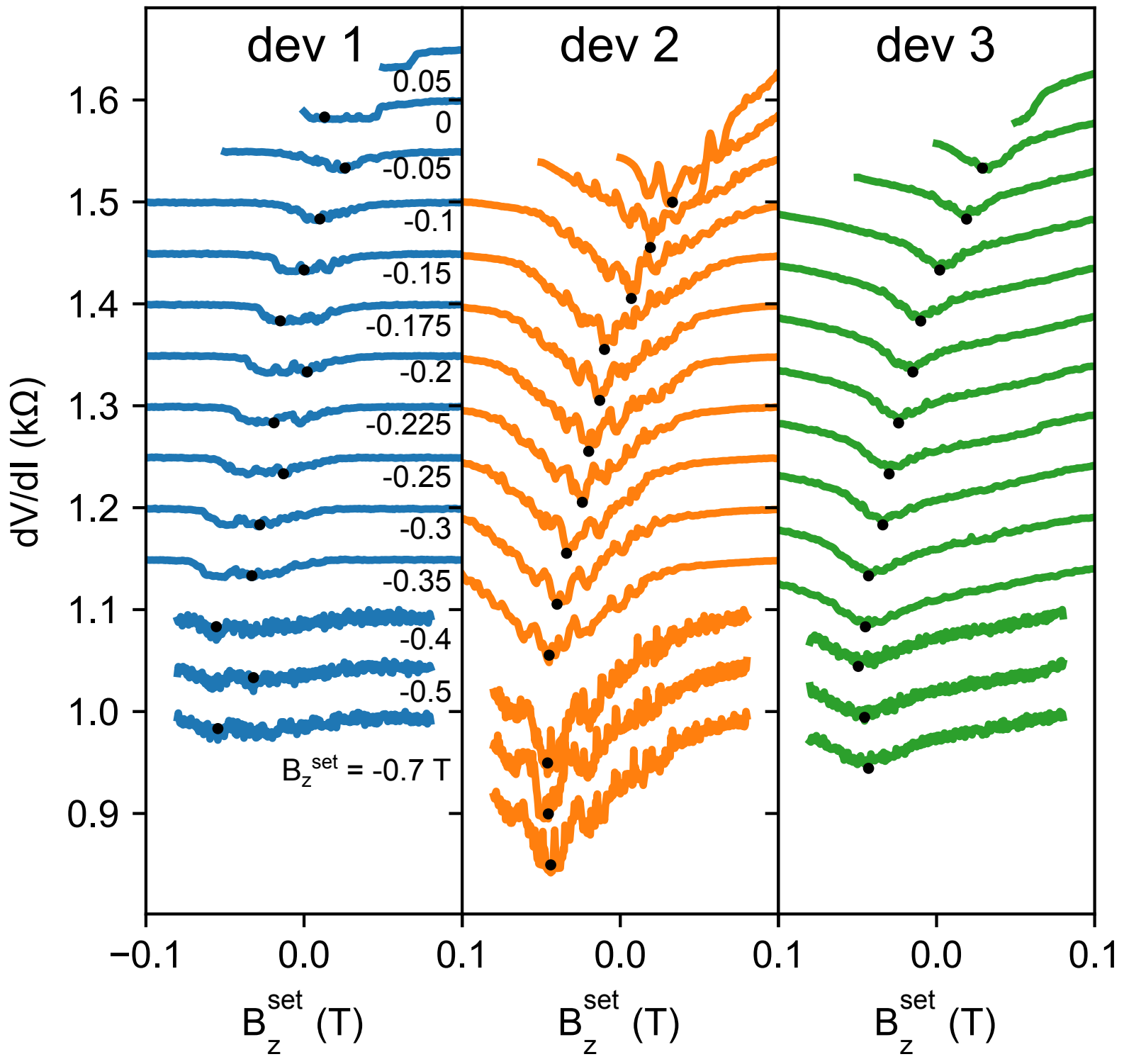


FIG. S3. Differential resistance as a function of magnetic field for different values of $B_z^{\rm set}$ for device 1, 2 and 3.

voltage. Consequently, the magnetic field range over which the superconducting state persists also increases. We attribute this enhancement of the critical current to the decrease in the normal-state resistance as additional charge carriers are accumulated in the 2DEG.

Figure S4(d) shows the switching current, extracted at the magnetic field indicated by the dashed lines in panels (a-c), together with the normal-state resistance ($R_n$), as a function of back-gate voltage. The corresponding $I_cR_n$ product remains approximately constant over the measured gate-voltage range, suggesting that the superconducting gap is approximately independent of back-gate voltage.

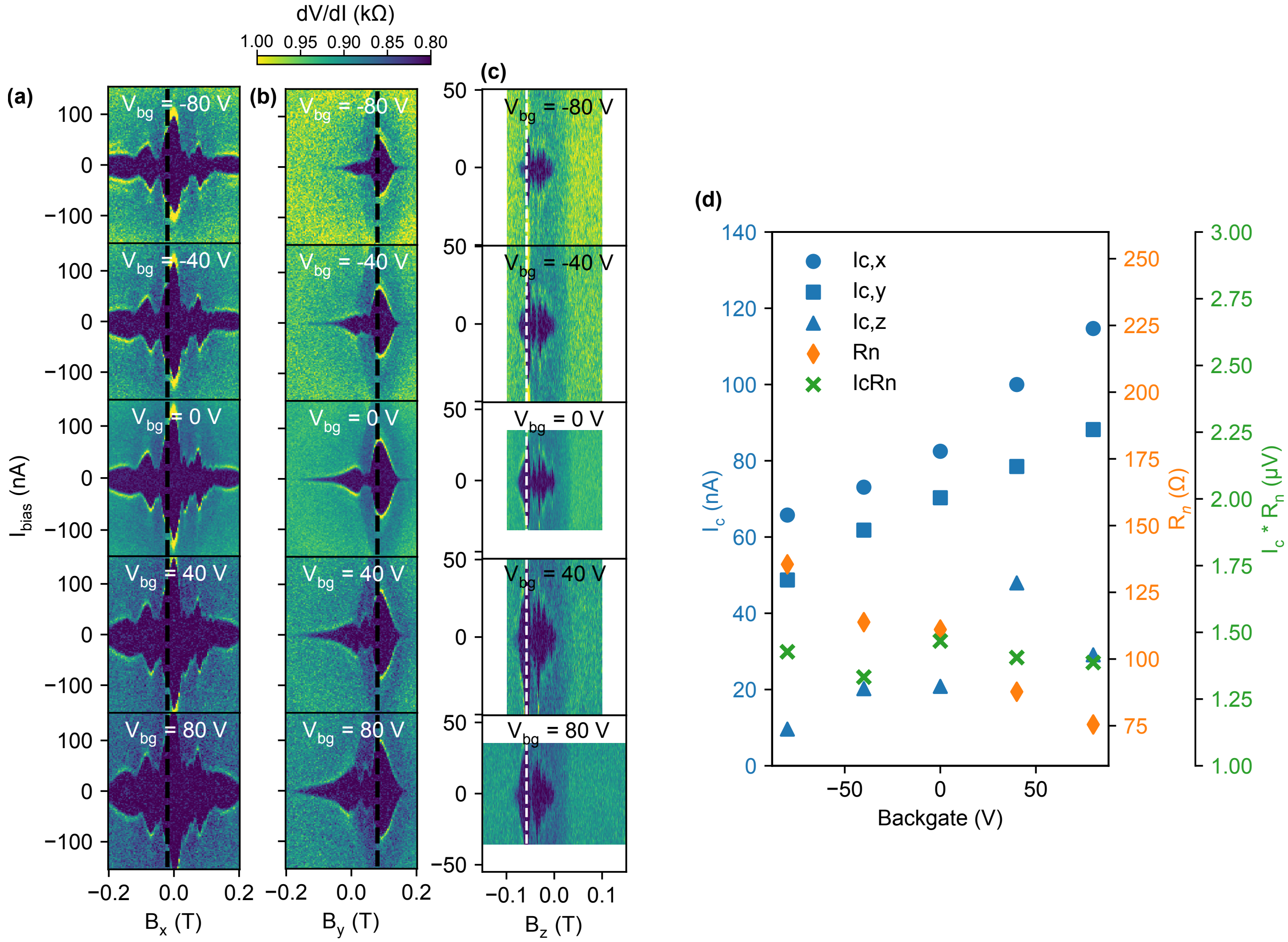


FIG. S4. (a), (b), (c), The differential resistance maps of device 1 as a function of current bias and magnetic field applied along the $x$- , $y$-, and $z$-directions, respectively, for electrostatic back-gate voltages ranging from $-80$ to 80 V. (d) Critical current in the x, y and z direction, normal resistance and $I_c R_n$ product as function backgate voltage.

## S5. COMSOL

To evaluate the magnetic stray fields originating from the nickel contacts, we performed finite element simulations using the commercially available software COMSOL Multiphysics® and its dedicated Magnetic Fields module.

The geometric dimensions of the simulated domains were chosen to match the physical size of the experimental samples. We used two different configurations: (1) To simulate the case where magnetization lies in-plane ($x$- and $y$-direction), the two ferromagnetic leads were modeled using a top-view 2D geometry consisting of two rectangles, which provides an excellent approximation for thin films. (2) For the magnetization pointing out-of-plane ($z$-direction), the cross-section in the $xz$-plane of the system was modeled. This is justified because the magnetic field exhibits translational invariance in $y$-direction far from the edges where the membrane is located. Both configurations are sketched in Fig. S5

Using these geometrical considerations, the spatial distribution and magnitude of the magnetic stray field within the membrane area were computed. Since no currents are involved ($\mathbf{J} = 0$), the governing physics reduces to magnetostatics. COMSOL solves Ampère's law in solids

$$\nabla \times \mathbf{H} = \mathbf{J} = 0 \tag{S1}$$

via the magnetic vector potential $\mathbf{A}$, defined by $\mathbf{B} = \nabla \times \mathbf{A}$, and with the constitutive relation

$$\mathbf{B} = \mu_0(\mathbf{H} + \mathbf{M}), \tag{S2}$$

where the magnetization $\mathbf{M}$ is zero in the surrounding air domain and non-zero within the nickel contacts, $\mathbf{H}$ the magnetic field strength and $\mathbf{B}$ the magnetic flux density, which we refer to as the 'magnetic field' in the main text. The resulting partial differential equation reads

$$\nabla \times \left( \frac{1}{\mu_0} \nabla \times \mathbf{A} - \mathbf{M} \right) = 0. \tag{S3}$$

For every direction, i.e. $x$, $y$ and $z$, we perform a parametric sweep of the magnetization, in order to obtain $\mathbf{B}_{\mathrm{2DEG}}(M_i)$, the effective stray field within the membrane as a function of the nickel contact magnetization (see main text Fig. 3(b) and Fig. 5(d),(e)).

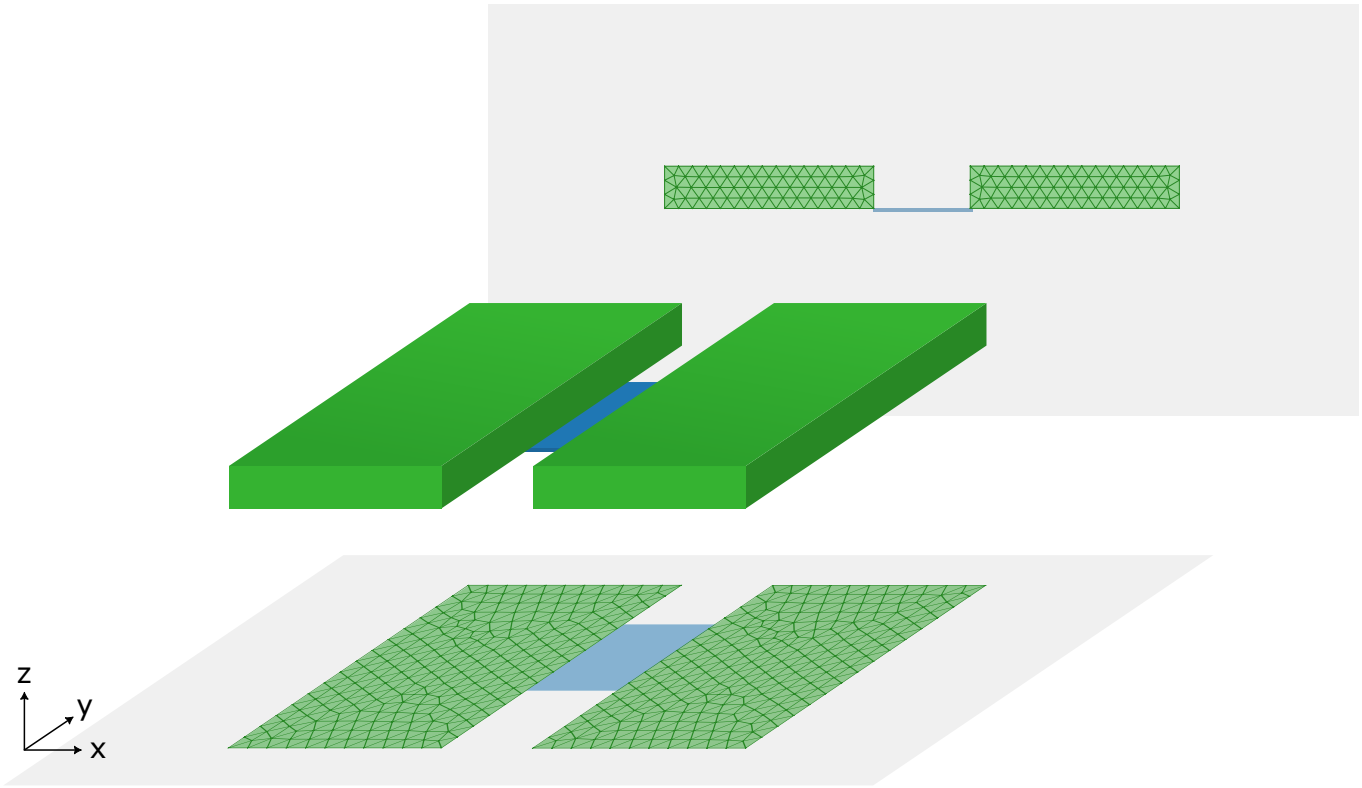


FIG. S5. Sketch of the geometry utilized for simulating the magnetic stray field of the nickel contacts (green color). Their shape is projected into the $xy$- and $xz$-plane. The area in which the stray field is averaged, corresponding to the membrane, is highlighted in blue.



\end{document}